# An 818-TOPS/W CSNR-31dB SQNR-45dB 10-bit Capacitor-Reconfiguring Computing-in-Memory Macro with Software-Analog Co-Design for Transformers

Kentaro Yoshioka

Keio University, Yokohama, Japan, Email: kyoshioka47@keio.jp

*Abstract*

Transformer inference requires high compute accuracy; achieving this using analog CIMs has been difficult due to inherent computational errors. To overcome this challenge, we propose a Capacitor-Reconfiguring CIM (CR-CIM) to realize high compute accuracy analog CIM with a 10-bit ADC attaining high-area/power efficiency. CR-CIM reconfigures its capacitor array to serve dual purposes: for computation and ADC conversion, achieving significant area savings. Furthermore, CR-CIMs eliminate signal attenuation by keeping the signal charge stationary during operation, leading to a 4x improvement in comparator energy efficiency. We also propose a software-analog co-design technique integrating majority voting into the 10-bit ADC to dynamically optimize the CIM noise performance based on the running layer to further save inference power.

Our CR-CIM achieves the highest compute-accuracy for analog CIMs, and the power efficiency of 818 TOPS/W is competitive with the state-of-the-art. Furthermore, the FoM considering SQNR and CSNR is 2.3x and 1.5x better than previous works, respectively. Vision Transformer (ViT) inference is achieved and realizes a highest CIFAR10 accuracy of 95.8% for analog CIMs.

## Introduction

CIM technologies have the potential to significantly reduce the energy requirements for DNN inference by minimizing data transfer. While many research efforts have been focused on improving the area and power efficiency of analog CIMs [1-6], there has been little focus on improving the *effective compute accuracy* of CIMs; CIM's analog noise and quantization noise are often not well-considered. Although relatively light networks such as CNNs can still achieve acceptable overall accuracy with low compute accuracy, significantly higher compute signal-to-noise ratio (CSNR)[1] is required for carrying out more complex applications such as Transformers, the next-generation algorithm for natural language processing, computer vision, and audio recognition, as illustrated in Fig.1 (A). Therefore, a more precise compute scheme for analog CIM is desired.

However, achieving high-CSNR is challenging in analog CIM due to the inherent analog computing errors. To effectively perform Transformer inference, analog CIMs must possess high levels of 1) linearity, 2) ADC resolution and 3) circuit noise resiliency, all while keeping circuit overhead to a minimum. Current [2] and time-based CIMs [3] are prone to mismatch and have difficulty achieving >8bit linearity levels. Charge-based CIMs [4, 5] have the potential to satisfy the linearity due to its high-matching property of capacitors but are impractical to scale up to the necessary 10-bit ADC resolution for Transformers, owing to the significant overheads in area and power consumption, as illustrated in Fig.1 (B). Due to such challenges, to the best of the author's knowledge, Transformer acceleration using analog CIMs have not been reported to date. While digital CIMs do not suffer from analog errors, their power efficiency can compete with analog CIMs only with advanced nodes such as 5nm [6]. To tackle the above issues, we propose a novel Capacitor-Reconfiguring analog CIM architecture that can reach high area/power efficiency and high-CSNR for CMOS-based edge and IoT applications.

## CR-CIM concept

Fig. 2 shows the proposed CR-CIM circuit which utilizes a reconfiguring capacitor array to serve dual purposes: 1) charge-based compute and 2)10-bit C-DAC for the 10-bit ADC operation.

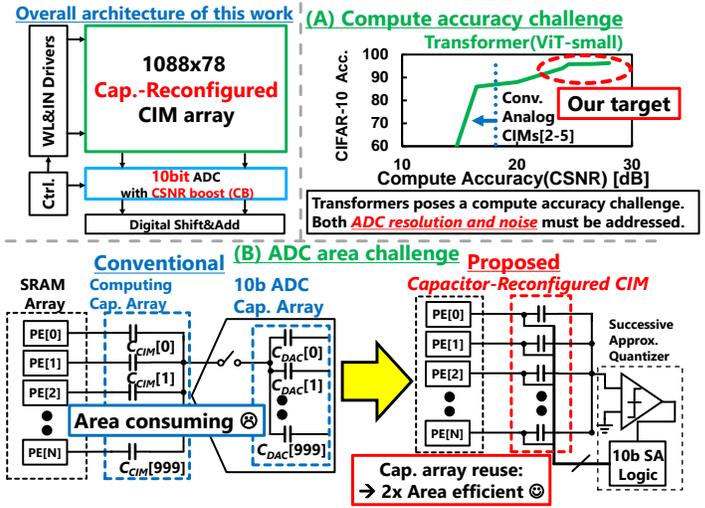

Fig.1 Challenges of high-accuracy CIMs.

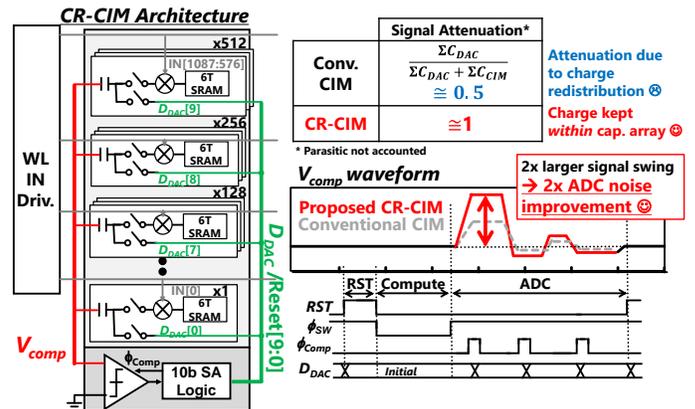

Fig.2 Proposed CR-CIM architecture.

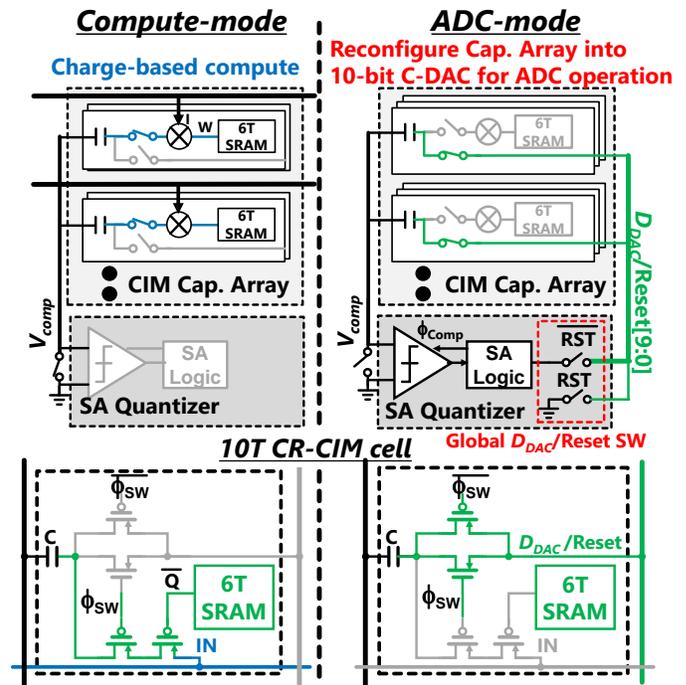

Fig.3 CR-CIM operation and cell schematic.

This realizes an analog CIM with high-resolution ADC while maintaining a significantly small area. During the computation phase, the product of the input (IN) and the weight stored in the 6T SRAM is fed into the CIM cell capacitor. In the ADC phase, the CIM capacitor is connected to the DAC feedback signal ($D_{DAC}$[9:0]) to enable reconfiguration to a C-DAC array. The signal $D_{DAC}$[9] is connected to 512 CIM cells, the signal $D_{DAC}$[8] is connected to 256 CIM cells, and so forth, forming a binary C-DAC array. In conventional CIMs, the charge obtained during computation is transferred to the ADC through charge redistribution, which leads to significant signal attenuation. CR-CIM addresses this issue by maintaining the signal charge within the capacitor array throughout the entire operation. The top plate of the CR-CIM's capacitor array is directly connected to the comparator, allowing for the implementation of successive approximation (SA) through $D_{DAC}$[9:0] feedback to achieve 10-bit AD conversion. CR-CIM achieves 2x larger signal swings in comparison to conventional CIMs, which is equivalent to 2x ADC noise improvements. While high-resolution SAR ADCs' power consumption is typically dominated by the low-noise comparators, the CR-CIM can significantly reduce its power consumption through relaxation of the comparator noise requirements.

CIMs require both compute and reset mechanisms, but CR-CIMs also necessitate the transmission of $D_{DAC}$ signals as an overhead. To minimize these overheads, we propose a $D_{DAC}$/reset sharing structure (Fig.3). The $D_{DAC}$ path is reused to reset the charge as well, eliminating the need for reset switches within the cell and enabling CR-CIM operation to be achieved with a small 10T cell. The shared $D_{DAC}$/Reset[9:0] node is connected to the reset voltage during reset (RST) and switched to $D_{DAC}$[9:0] during the ADC phase by a global switch. CIM cell capacitor is realized with a custom 1.5fF fringe capacitor and confirmed to exhibit 10-bit linearity in post-layout simulations. CR-CIM cell area was 2.3um² using 65nm logic rules, which is approximately 2x larger than a typical 6T SRAM cell.

Fig. 4 illustrates our strategy for improving the inference efficiency of the Transformer through software-analog co-design (SAC). The Transformer can be broken down into two layers: an Attention block that mixes feature vectors of image patches, and a vector-wise MLP block. We utilize the following observation: the required CSNR of the Attention layer is 10dB lower than that of the MLP layer. We propose a CSNR boost (CB) technique that allows for a trade-off between readout accuracy and power. By adaptively activating CB based on the running Transformer block, inference efficiency can be improved up to 2.1x. When CB is enabled, 6x majority voting (MV) is applied to the last 3 SA comparisons and increases the CSNR by 5.5dB, but in turn of power and conversion time overhead by 1.9x and 2.5x, respectively. Although MV is a common noise reduction technique for SAR ADCs, this is the first SAC that adaptively utilize MV reflecting the running DNN layer. Additionally, while it is challenging to implement a low-noise comparator with a narrow array pitch, CB relaxes the noise requirement and simultaneously eases layout practicality.

### Measurement results

The CR-CIM was prototyped as a 1088x78 array in 65nm CMOS. Fig. 5 summarizes the measured column characteristics of CR-CIM. Measured results show good CIM transfer characteristics with INL error within <2 LSBs, despite the challenging 10-bit readout. With CB, the noise is 0.58 LSB on average for all codes and increases by 2x when CB is disabled. SQNR [4] and CSNR [1], in which circuit noise is considered, are 45 dB and 31 dB, respectively. These metrics are 23 dB and 14 dB better than the previously reported charge-based CIMs [4,5], showing that CR-CIM and CB techniques lead to high compute accuracy.

Fig. 6 summarizes the performance of the CR-CIM. To demonstrate the acceleration of the Transformer, CIFAR-10 accuracy was confirmed using a Vision Transformer (ViT-small) network with 12 stacked transformer layers. CIM computes the Linear layers, where the MLP layer was run w/CB 6-bit precision, and the Attention layer was run wo/CB with 4-bit precision for both input and weights, respectively. As a result, 95.8% TOP-1 accuracy is achieved, which is comparable to the ideal inference (96.8%). Simulation analysis showed that with the proposed SAC and bit precision optimization, the Transformer inference efficiency is improved by 2.1x. Comparing with SQNR FoM and CSNR FoM, which takes SQNR and CSNR into account, this work is 2.3x and 1.5x higher than previous analog CIMs, respectively.

*Acknowledgements* This work was supported by JST CREST program JPMJCR21D2 and Futaba Foundation.

### References
[1] S. Gonugondla, et al, "Fundamental Limits on the Precision of In-memory Architectures," *ICCAD*, 2020.
[2] Q. Dong, et al, "A 351TOPS/W and 372.4GOPS Compute-in-Memory SRAM Macro in 7nm FinFET CMOS for Machine-Learning Applications," *ISSCC*, 2020.
[3] P. Wu et al, "A 28nm 1Mb Time-Domain Computing-in-Memory 6T-SRAM Macro with a 6.6ns Latency, 1241GOPS and 37.01TOPS/W for 8b-MAC Operations for Edge-AI Devices," *ISSCC*, 2022.
[4] H. Jia, et al, "A Programmable Heterogeneous Microprocessor Based on Bit-Scalable In-Memory Computing," *IEEE JSSC*, 55(9), 2609-2621, 2020.
[5] J. Lee, et al, "A 13.7 TFLOPS/W floating-point DNN processor using heterogeneous computing architecture with exponent-computing-in-memory," *Symp. VLSI 2021*.
[6] H. Fujiwara et al, "A 5-nm 254-TOPS/W 221-TOPS/mm2 Fully-Digital Computing in-Memory Macro Supporting Wide-Range Dynamic-Voltage Frequency Scaling," *ISSCC* 2022.

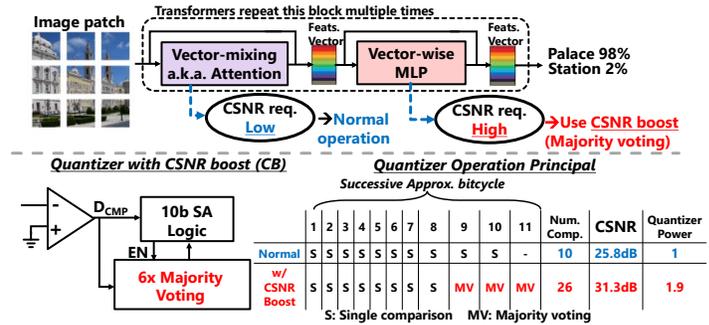

Fig.4 Transformer Software-Analog Co-Design (SAC).

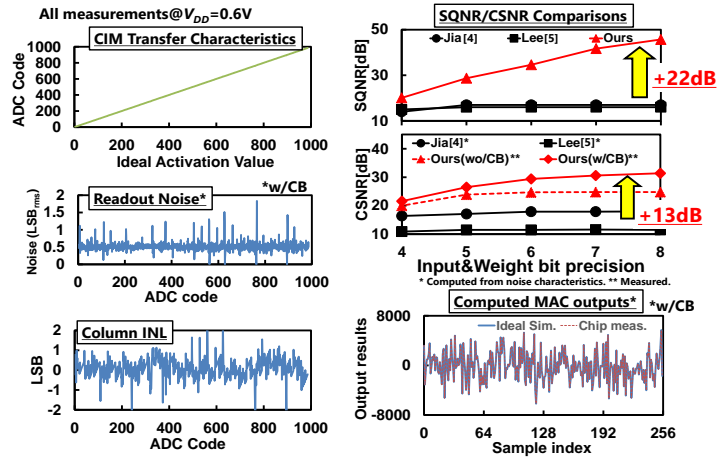

Fig.5 Measured CR-CIM column characteristics.

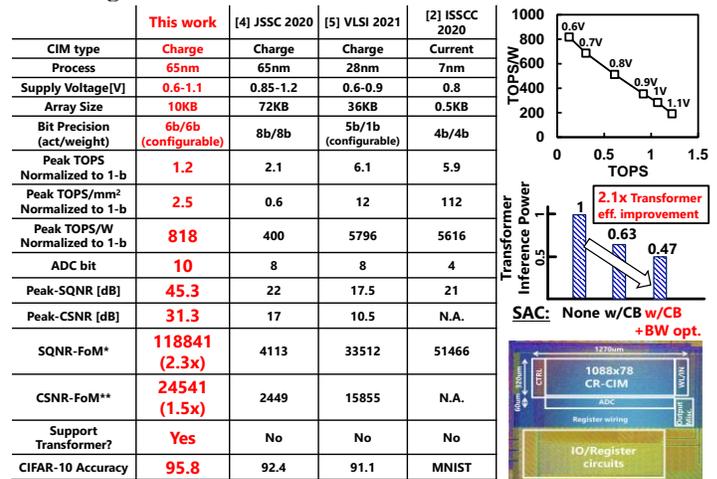

| | This work | [4] JSSC 2020 | [5] VLSI 2021 | [2] ISSCC 2020 |
|---|---|---|---|---|
| CIM type | Charge | Charge | Charge | Current |
| Process | 65nm | 65nm | 28nm | 7nm |
| Supply Voltage[V] | 0.6-1.1 | 0.85-1.2 | 0.6-0.9 | 0.8 |
| Array Size | 10KB | 72KB | 36KB | 0.5KB |
| Bit Precision (act/weight) | 6b/6b (configurable) | 8b/8b | 5b/1b (configurable) | 4b/4b |
| Peak TOPS Normalized to 1-b | 1.2 | 2.1 | 6.1 | 5.9 |
| Peak TOPS/mm² Normalized to 1-b | 2.5 | 0.6 | 12 | 112 |
| Peak TOPS/W Normalized to 1-b | 818 | 400 | 5796 | 5616 |
| ADC bit | 10 | 8 | 8 | 4 |
| Peak-SQNR [dB] | 45.3 | 22 | 17.5 | 21 |
| Peak-CSNR [dB] | 31.3 | 17 | 10.5 | N.A. |
| SQNR-FoM* | 118841 (2.3x) | 4113 | 33512 | 51466 |
| CSNR-FoM** | 24541 (1.5x) | 2449 | 15855 | N.A. |
| Support Transformer? | Yes | No | No | No |
| CIFAR-10 Accuracy | 95.8 | 92.4 | 91.1 | MNIST |

* SQNR-FoM=TOPS/W*2^SQNRbit  ** CSNR-FoM=TOPS/W*2^CSNRbit
*** SQNRbit = (SQNR[dB]-1.76)*6.02   CSNRbit = (CSNR[dB]-1.76)*6.02

Fig.6 Performance summary.